\documentclass[aps,twocolumn,showpacs,nofootinbib,showkeys,superscriptaddress,preprintnumbers,longbibliography,10pt]{revtex4-2}
\usepackage[utf8]{inputenc}

\usepackage{latexsym}
\usepackage{epsfig}
\usepackage{graphicx,subfigure}
\usepackage[normalem]{ulem}
\usepackage{color}
\usepackage{xcolor}
\usepackage{multirow}
\usepackage{hyperref}
\usepackage{numprint}
\usepackage{eucal}
\usepackage{amsmath}
\usepackage{amssymb}
\usepackage{amsmath,amssymb} 
\usepackage{wasysym}

\DeclareFontFamily{U}{mathb}{\hyphenchar\font45}
\DeclareFontShape{U}{mathb}{m}{n}{
      <5> <6> <7> <8> <9> <10> gen * mathb
      <10.95> mathb10 <12> <14.4> <17.28> <20.74> <24.88> mathb12
      }{}

\definecolor{blue}{rgb}{0,120,250}

\usepackage{xcolor}

\newcommand{\ssout}[1]{}
\begin{document}

\def\ga{\mathrel{\raise.3ex\hbox{$>$\kern-.75em\lower1ex\hbox{$\sim$}}}}
\def\la{\mathrel{\raise.3ex\hbox{$<$\kern-.75em\lower1ex\hbox{$\sim$}}}}

\def\be{\begin{equation}}
\def\ee{\end{equation}}
\def\bea{\begin{eqnarray}}
\def\eea{\end{eqnarray}}

\def\betap{\tilde\beta}
\def\del{\delta_{\rm PBH}^{\rm local}}
\def\Msun{M_\odot}

\newcommand{\dd}{\mathrm{d}} 
\newcommand{\Mpl}{M_P} 
\newcommand{\mpl}{m_\mathrm{pl}} 

\newcommand{\CHECK}[1]{{\color{red}~\textsf{#1}}}

\title{Primordial Black Holes without fine-tuning\\ from a light stochastic spectator field } 

\author{Ioanna Stamou}
\affiliation{Service de Physique Th\'eorique, Universit\'e Libre de Bruxelles (ULB), Boulevard du Triomphe, CP225, B-1050 Brussels, Belgium}

\author{Sebastien Clesse}
\affiliation{Service de Physique Th\'eorique, Universit\'e Libre de Bruxelles (ULB), Boulevard du Triomphe, CP225, B-1050 Brussels, Belgium}
 
\begin{abstract} 
We investigate a mechanism of primordial black hole (PBH) formation that avoids any dependence on specific inflationary features or exotic physics.
In this scenario, the required large curvature fluctuations leading to PBH formation are generated after inflation by the quantum fluctuations of a light stochastic spectator field during inflation, when this field transiently dominates the energy density.  We calculate the dynamics of such a spectator field during and after inflation, the distribution of induced curvature perturbations and their non-Gaussian tails leading to the copious production of PBHs.  
For a plateau-like potential, this scenario produces an extended PBH mass distribution with a peak at the solar-mass scale when one takes into account the effects of the thermal history. What is remarkable in this scenario is the absence of parameter fine-tuning.  Instead, it invokes an anthropic selection over all the realizations of PBH abundances predicted by the field stochasticity.  
This scenario offers a novel perspective for the formation of PBHs with minimal ingredients and without the need of fine-tuning.  It is amenable to observational tests, notably with the gravitational-wave observations of black hole mergers and of a background at nanoHertz frequency, as recently observed by pulsar timing arrays.

\end{abstract}

\maketitle


\section{Introduction} 

Taking advantage of the absence of detection of new particles such a weakly interacting massive particles, in accelerators and in direct and indirect detection experiments, primordial black holes (PBHs) are nowadays considered as one leading candidate to explain the dark matter in the Universe.   Contrary to dark matter particles, the existence of PBHs is supported by a series of observations, reviewed in~\cite{Clesse:2017bsw,Carr:2019kxo,Cappelluti:2021usg,Carr:2023tpt} 
and including the gravitational waves (GW) from compact binary coalescences observed by the Ligo/VIRGO/Kagra (LVK) collaboration~\cite{Abbott:2016blz,LIGOScientific:2016dsl,LIGOScientific:2018mvr,LIGOScientific:2020ibl,LIGOScientific:2021usb,LIGOScientific:2021djp}, a GW background at nanoHertz frequency detected with pulsar timing arrays (PTA)~\cite{NANOGrav:2023hvm,NANOGrav:2023hfp,Antoniadis:2023ott,Zic:2023gta,Reardon:2023gzh,Xu:2023wog}, the size and mass-to-light ratio of ultra-faint dwarf galaxies, several microlensing candidates, spatial correlations in source-subtracted cosmic infrared and X-ray backgrounds, the existence of supermassive black holes at high redshifts (see~\cite{Carr:2023tpt} and references therein).  
These observational clues are however not unambiguous and could have other astrophysical origins.  

In addition, there are also numerous constraints on the abundance of PBHs, see e.g.~\cite{Carr:2020xqk}
 for a recent review, sometimes in apparent conflict with some of those hints.   Furthermore, it is worth noticing that any observational evidence or constraint is still subject to large uncertainties or model dependence.  It is therefore very difficult to prove the existence of PBHs and, if they exist, to infer their total contribution to the dark matter.   There is so far only one almost unambiguous way to prove the existence of PBHs that is accessible with the current generation of instruments:  detecting a subsolar-mass black hole in a compact binary coalescence.   
Recently a few intriguing subsolar-mass triggers have been reported in GW observations~\cite{Phukon:2021cus,LIGOScientific:2022hai}.  For instance, SSM170401 prefers a subsolar-mass black hole secondary component if interpreted as a GW signal~\cite{Morras:2023jvb}.  Overall, the search for PBHs and their properties is a very active and exciting area of research, with many implications for our understanding of the nature of dark matter and of the physics at play in the early Universe.

PBHs are thought to have formed from the collapse of regions of high density contrast in the early Universe.   An important criticism of the  majority of PBH scenarios comes from the difficulty to produce them without invoking strong parameter fine-tuning~\cite{Cole:2023wyx} and specific models of the early Universe, such as transient inflationary features in the primordial power spectrum, during reheating or new phase transitions (see e.g.~\cite{Ozsoy:2023ryl,Sasaki:2018dmp,Khlopov:2008qy} for a review).  For instance, the mechanism of PBH formation may involve the amplification of quantum fluctuations during inflation.  A lot of PBH models rely on this idea but they require a strong enhancement of the primordial power spectrum at small scales.   
Such a feature is not natural in the vast majority of single-field slow-roll inflation models.  It typically requires an extremely flat region of the scalar field potential over a tiny field range, leading to a so-called transient phase of \textit{ultra-slow-roll}.  In addition, in most models the abundance of PBHs depends exponentially on the amplitude of those fluctuations, leading to an additional layer of fine-tuning for the model parameters~\cite{Cole:2023wyx,Stamou:2021qdk}.

 In this work we explore a mechanism of PBH production based on a light quantum stochastic spectator scalar field during inflation.  By definition, a spectator field is a hypothetical scalar field, not involved in the inflationary expansion of the early Universe.  Inflation therefore does not play a direct role in the PBH production, and vice-versa.  Because the field is very light, the exact shape of its potential is also irrelevant for the dynamics of its quantum fluctuations during inflation, which adds to the genericity of the scenario and allows PBH formation with relatively minimal assumptions and no strong dependence on potential parameters.   It can have a variety of origins and properties, spectator fields being generic in various high-energy frameworks like supersymmetry, supergravity, grand unified theories, string theory, extra-dimension models, etc.   The best example of a light spectator field during inflation is the Broug-Englert-Higgs (BEH) field.  We consider this specific case in a companion paper and focus here on the more general case.  Contrary to most other PBH models based on extra spectator fields~\cite{Carr:2017edp,Garcia:2020mwi,Cai:2021wzd,Maeso:2021xvl,Cable:2023lca}, the curvature fluctuations at the origin of PBHs are not produced during inflation in our scenario, but after, in the subsequent matter or radiation era.  This happens when the field starts to dominate the energy density of the Universe and drive the expansion, in such a way that small fluctuations of the field can be converted into large but still super-horizon curvature fluctuations, as in the so-called \textit{curvaton} scenario~\cite{Lyth:2002my} but with the addition that curvature fluctuations are highly non-Gaussian.  More precisely, in some rare regions where the spectator field lies in a sufficiently flat region of its potential, one gets a short extra phase of expansion, i.e. a curvature fluctuation.  PBHs are only produced later, when these fluctuations re-enter inside the Hubble radius and collapse gravitationally.   

 This  mechanism was proposed in \cite{Carr:2019hud} by one of us, but using simple assumptions and estimates.  Other authors also proposed more specific models based on a spectator field and sometimes qualitatively similar scenarios~\cite{Passaglia:2019ueo,Meng:2022low,Gow:2023zzp,Shinohara:2021psq,Pi:2022ysn,Shinohara:2023wjd,Ota:2022hvh}. In this work, we refine the analysis of the stochastic dynamics during inflation and improve the calculation of the resulting PBH abundance.  We compute the exact field and expansion dynamics leading to the generation of curvature fluctuations, for realistic example models, considering both the field and the radiation or matter content in the Universe at this epoch, instead of assuming simple slow-roll conditions.  We show that this affects the conditions for the realisation of this scenario and modifies the PBH mass distribution.  Furthermore, we estimate the contribution of curvature fluctuations from  the spectator field to the total primordial power spectrum, on cosmological scales and on smaller scales, using the stochastic $\delta N$ formalism \cite{Lyth:2005fi,Langlois:2008vk,Sugiyama:2012tj}.  We find the conditions under which they are subdominant on cosmological scales.  The statistics of these  fluctuations is also investigated and we find that on all scales it exhibits a non-Gaussian tail, which triggers the formation of PBHs. 
 Finally, we include the effect of the QCD cross-over transition on the critical overdensity threshold, leading to specific features in distribution of stellar-mass PBHs that can be tested with GW observations of compact binary coalescences.

The layout of this paper is as follows. In Section \ref{Stochastic spectator during inflation} we introduce and solve the stochastic dynamics of a light spectator field during inflation. In Section \ref{Two example models}, two examples of scalar field potentials are introduced, one successfull (plateau potential) and one unsuccessfull (small-field potential). In Section \ref{The Dynamics of the Field after inflation}, we solve the exact dynamics of the field after inflation when it transiently dominates the Universe.  The resulting production of curvature fluctuations and their statistics are analyzed in Section\ref{Production of curvature fluctuations after inflation}, where the condition that they are subdominant on large cosmological scales is imposed.  In Section \ref{Primordial Black Hole formation}, we compute the abundance of PBHs and their mass distribution in several cases, including the effects of the QCD transition.  We then explain and discuss why there is no fine-tuning issue in our scenario in Section \ref{fine-tuning}.  We try to quantify the implications of this property in terms of Bayes factor when our model is compared to some other PBH scenarios.  Finally, in Section \ref{Discussion and conclusion}, we discuss our results and we present our conclusions, as well as the perspepectives of our work, with a focus on how to distinguish the model observationnally.

\section{Quantum stochastic dynamics during inflation}
\label{Stochastic spectator during inflation}

In the considered scenario, the production of PBHs results from the stochastic quantum fluctuations of a light spectator field during cosmic inflation.  In different Universe patches that are comparable to the size of our observable Universe, {referred as Hubble-sized patches}, the field has acquired different mean values, different fluctuation statistics, {leading to} different PBH abundances.  Since inflation can generically lead to much more than 60 e-folds of expansion, there are today so many of these Hubble-sized patches that there is necessarily one associated to a given PBH abundance.   In order to calculate the abundance of PBHs with a given mass in a given patch, the first step is thus to solve the stochastic dynamics of the field during inflation, which is the goal of this section.

The inflationary dynamics is commonly described using the so-called Hubble-flow (slow-roll) functions denoted by $\mathrm{\epsilon_{1,2,3}}$ and defined as follows, 
\begin{equation}
\epsilon_1 \equiv -\frac{{\rm d} \ln H}{{\rm d}N},\quad \epsilon_2 \equiv \frac{{\rm d} \ln \epsilon_1}{{\rm d}N},\quad \epsilon_3 \equiv \frac{{\rm d} \ln|\epsilon_2|}{{\rm d}N},
\end{equation}
where  $H$ is the Hubble-Lema\^itre expansion rate and $N$ denotes the number of e-folds realized during inflation.  We arbitrarily fix $N=0$ when our Hubble-sized patch exited the Hubble-radius during inflation.  These slow-roll functions can be reconstructed from the amplitude and scale-dependence of the (scalar) primordial perturbations that seeded the observed large-scale structures and cosmic microwave background anisotropies.  This primordial power spectrum is usually parameterized by its amplitude ${A_{\rm s}}$ and spectral index $n_{\rm s}$ at the pivot scale $k_* = 0.05 {\, \rm Mpc}^{-1}$. {In the context of single-field slow-roll inflation, its amplitude is given by} 
\begin{equation}
A_{\rm s}=2.1 \times 10^{-9}\simeq \frac{H^2_{*}}{8\pi^2 \epsilon_1 M_{\rm P}^2}~,
\label{eq:amplitude}
\end{equation}
and the spectral index is 
\begin{equation}
n_{\rm s} =0.9649 \pm	0.0042\simeq	 1-2\epsilon_{1*} -\epsilon_{2*}~,
\label{eq.ns}
\end{equation}
according to the latest observations by Planck~\cite{Planck:2018jri}, where $M_\mathrm{P}$ is the reduced Planck mass and a star subscript means that a quantity is evaluated at the time when the {comoving} pivot scale $k_*$ exited the Hubble radius, for instance $H_*$ is the expansion rate when $k_* = a H_*$.  
For the inflation dynamics, we consider two {phenomenological} models that are consistent with the above-mentioned constraints and with the current limit on the tensor-to-scalar ratio $r \simeq 16 \epsilon_{1*} \lesssim 0.05$~\cite{Planck:2018jri}, reported in Table~\ref{tab:epsi}.  For simplicity we assumed that $\epsilon_3 = 0$ but our results can easily {be} generalized to other models or to refined inflationary predictions.  It is important to note that the exact shape of the PBH mass distribution will {marginally} depend on the assumed inflation dynamics, but that the process of PBH formation itself remains generic and does not require any specific inflationary scenario.  
\begin{table}[!h]
\begin{center}
\begin{tabular}{||c| c |c |c|c||} 
 \hline
 Model & $\mathrm{\epsilon_{1*}}$ & $\mathrm{\epsilon_{2*}}$ & $\mathrm{H_{*} [M_P]}$ & r \\ [0.5ex] 
 \hline\hline
1&0.00507 &0.0207& 2.9 $\times10^{-5}$ &0.08115 \\
 \hline
2 &  0.00020 & 0.0351 &5.8$\times10^{-6}$ & 0.00325\\
 \hline
\end{tabular}
\caption{\label{tab:epsi} Hubble rate and Hubble-flow functions for the pivot scale $k_* = 0.05 \, {\rm Mpc}^{-1}$ for the two illustrative benchmark models of inflation considered in this paper (without losing generality on the mechanism of PBH formation). }
\end{center}
\end{table}

The first model is representative of an inflaton potential (close to) linear leading to $\epsilon_2 \simeq 4 \epsilon_1$ and to a still acceptable value (despite disfavored) of the tensor-to-scalar ratio, at the limit of being detected.  The second model corresponds to the best and simplest inflationary model after Planck~\cite{Martin:2013nzq}, Higgs or Starobinsky inflation, and is representative of all the models where the spectral index value is saturated by the $\epsilon_{2*}$ parameter, while $\epsilon_1$ remains sufficiently small during inflation for not playing a significant role in the stochastic dynamics of the spectator field, {as later shown}.  These properties are typical of plateau-like potentials favored after Planck observations~\cite{Martin:2013nzq}. These models, with small variations, were also considered in the previous analysis of~\cite{Carr:2019hud}.

Let us now focus on the evolution of the quantum fluctuations of a light spectator field $\psi$  during inflation.  Each e-fold of expansion, the {amplitude of its quantum fluctuations} in a Hubble-sized region is of order $H/2\pi$. We denote {$\delta \psi_{\rm in}(x,N_{\rm inf})\equiv \psi_{\rm in}(x,N_{\rm inf})- \psi_{\rm out}(x,N_{\rm inf}-1)$} the variation of the mean {value of $\psi$} in a Hubble-sized {\textit{region}}\footnote{{Note the difference of terminology between a \textit{Hubble-sized region} denoting a spatial region of the Universe of size comparable to the Hubble radius at a given time, and a \textit{Hubble-sized patch} that corresponds to a spatial region of size comparable to the observable Universe.}} at e-fold time $N_{\rm inf}$ {centered on} the position $x$ with respect to the mean field value in the outer encompassing Hubble sized region at e-fold time $N_{\rm inf}-1$, denoted $\psi_{\rm out}(x,N_{\rm inf}-1)$.  The mean field value in the much larger Hubble-sized patch corresponding to our observable Universe is denoted $\langle \psi \rangle$.  Finally, $\delta \psi_{\rm out} \equiv \psi_{\rm out}(x,N_{\rm inf}-1) - \langle \psi \rangle $ refers to the field fluctuation between that outer region and the mean value in our observable Universe.  {These different quantities are illustrated by a sketch of our coarse-grained model shown in Fig.~\ref{fig:scheme}.}

The Fokker-Planck equation can be used for the  calculation of the probability distribution of the field during inflation, taking into account its quantum stochastic fluctuations~\cite{Hardwick:2017fjo,Carr:2019hud}.   This equation involves the drift and diffusion coefficients of the field, which are determined by the properties of the potential and the background cosmology.   For a very light spectator field with a mass $m \ll H_{\rm inf}$, however, the spectator field fluctuations remain independent of its potential.  The probability distribution can be used to calculate statistical quantities, such as the mean and variance of the field, and to study the behavior of the field in different regions of the Universe.  

{The quantum fluctuations of $\psi$ produced during one e-fold in a Hubble-sized region are Gaussian and} their variance is $H/2 \pi$.  Since $H$ evolves during inflation, driven by the slow-roll functions, if $\mathrm{\epsilon_2}$ is assumed to be constant, one has
\begin{equation} 
    \langle \delta \psi_{\rm in}^2(N_{\rm inf}) \rangle \simeq\frac{{H}^2_{*}}{4 \pi^2 } \exp\left\{ -2\frac{\epsilon_{1*}}{\epsilon_{2*}} \left[ e^{\epsilon_{2*} (N_{\rm inf} - N_*)} - 1 \right]\right\},
\label{eq:dhin}
\end{equation}
{where $N_* = \ln (k_* / H_0)  \simeq 5$.  If one is instead interested by the variance of the field fluctuations in a Hubble-sized patch, 
it is obtained through
\begin{equation}
 \langle \delta \psi^2(N_{\rm inf}) \rangle \simeq \int_0^{N} \frac{H(N)^2}{4 \pi^2} {\rm d} N.
\end{equation}
As long as $N_{\rm inf}-N_* \ll 1/\epsilon_2 \sim 50$ that is a valid hypothesis for stellar-mass PBHs for which $N_{\rm inf} - N_*\simeq 20$, one can expand the second exponential in $H^2(N)$ given by Eq.~\ref{eq:dhin} as $\exp[\epsilon_{2*} (N-N_*)] \simeq 1+ \epsilon_{2*} (N-N_*) $ and therefore one gets
{
\begin{eqnarray}
    \langle \delta \psi_{\rm out}^2(N_{\rm inf}) \rangle &\simeq & \frac{{H}^2_{*}}{8 \pi^2 \epsilon_{1*}}\left[ 1-\exp({-2\epsilon_{1*} (N_{\rm inf}-N_*)})\right]~, \nonumber \\
    & \simeq & \frac{H_*^2 (N_{\rm inf} - N_*)}{4 \pi^2} 
\label{eq:dhout}
\end{eqnarray}
It is worth noticing that the variance of field fluctuations grows linearly with the number of e-folds.  This property will be important to avoid large non-Gaussianities on cosmological scales and allow PBH formation on much smaller scales.  Those distributions will be used to compute the associated curvature fluctuations produced after inflation.


\begin{figure}[h!]
\centering
\includegraphics[width=80mm]{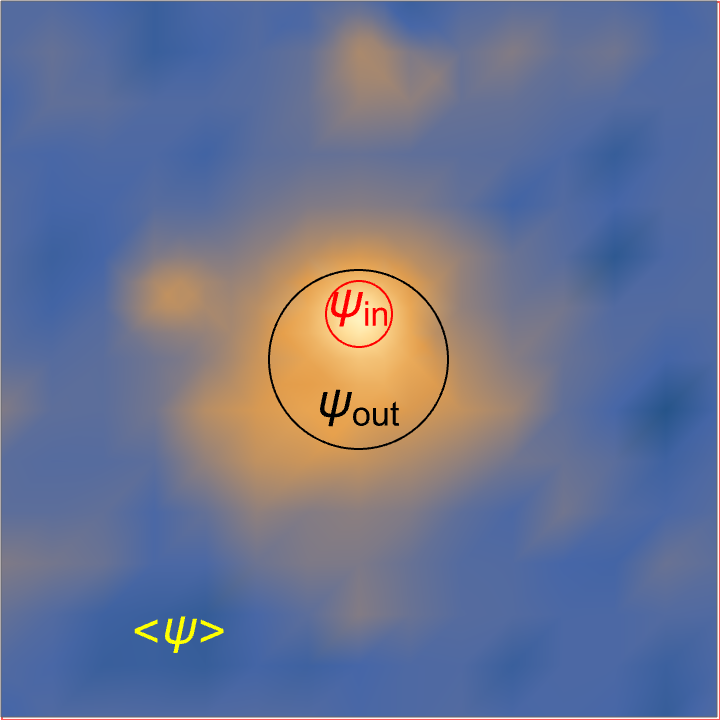}
\caption{Schematic representation of the spectator field fluctuations in a two-dimensional real-space slice, showing a rare fluctuation leading to a field value $\psi_{\rm out}$ that exited the Hubble radius at the e-fold time $N_{\rm inf} - 1$ and another smaller fluctuation on top of it with a field value $\psi_{\rm in}$ that became super-Hubble at time $N_{\rm inf}$.  If the yellow color represents field values in a sufficiently flat region of the potential, the inner fluctuation can lead to a large curvature fluctuation collapsing into a PBH when it re-enters the Hubble raidus.   }
\label{fig:scheme}
\end{figure}

\section{Example models}
\label{Two example models}
In this section, we present {two illustrative potentials for the light stochastic spectator field that can lead to the production of PBHs.  The first one is an archetype of a plateau-like potential, the second of small-field inflation.  However, as shown later, even if both can lead to the PBH production, only the former one does not overproduce at the same time curvature fluctuations on large cosmological scales, which is ruled out by CMB observations. 
}

{\textbf{Case 1:} The first considered potential has the  form:}
\begin{equation}
V(\psi)= \Lambda^4 \left( 1-  \exp \left[ -\frac{\psi}{M}\right]\right)
\label{eq:pot1}
\end{equation}
where  $\mathrm{\Lambda}$ {and} $\mathrm{M}$ are {two} parameters. This exponential potential has a plateau at large values of the field when { $\psi > M$}. 
In Fig.~\ref{f1} we depict the potential for {two relevant} choices of the parameters. {  With such a potential, only the plateau region is sufficiently flat to lead to extra e-folds of expansion.   In order to get sufficiently small curvature fluctuations in most regions of our Hubble-sized patch, we will have to consider patches where $\langle \psi \rangle$ does not lie in the plateau, together with a value of $M$ that is sufficiently small (but not too small) for quantum fluctuations of $\psi $ during inflation to lead to a subdominant fraction of Hubble-sized regions with $\psi(x,N_{\rm inf})$ in the plateau region, as required for having PBH formation.   Finally, it is worth noticing that the $\Lambda$ parameter does not influence the filed dynamics when expressed in e-fold time, one can therefore simply requires that $\Lambda$ is below the energy scale of inflation and above the QCD scale. }  
\begin{figure}[h!]
\centering
\includegraphics[width=80mm]{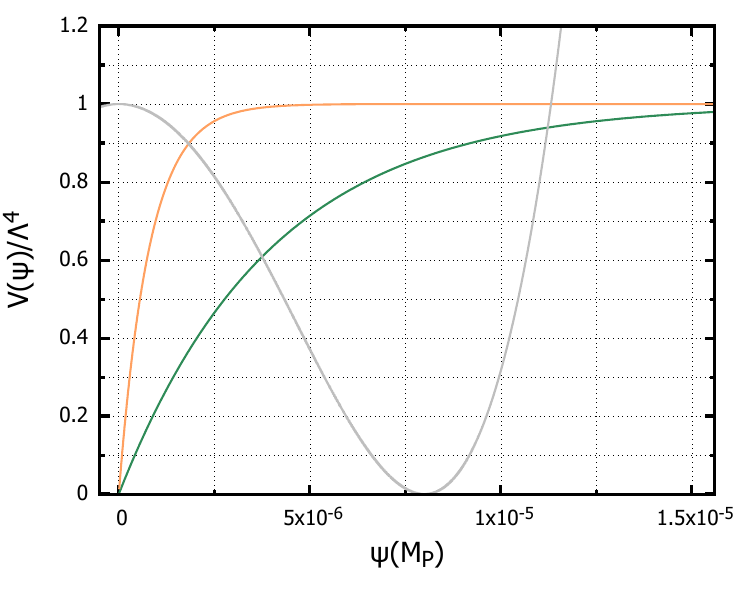}
\caption{The spectator field plateau-like potential of Eq.~(\ref{eq:pot1}), referred as Case 1.  The green line corresponds to the parameter ${M=4\times 10^{-6} M_{\rm P}}$ that will be used in combination with the inflation Model 1.  The orange line is obtained for ${M=8\times 10^{-7} M_{\rm P}}$ and is used for the inflation Model 2.  The gray line corresponds to the small-field spectator field potential of Eq.~(\ref{eq:potential_example2}), with ${M=8 \times 10^{-6}}$ and is considered as a representative unsuccessful model. }
\label{f1}
\end{figure}

{\textbf{Case 2:} The second potential we consider has} the following form,
 \begin{equation}
V(\psi)  = \Lambda^4	\left( 1 - \frac{\psi^2}{M^2}\right)^2~.
\label{eq:potential_example2}
\end{equation}
{It is a double well potential, with two minima at $\psi = \pm M$ while it exhibits a maximum and so a flat region at $\psi = 0$, where extra e-folds could be realised.  It grows like $V \propto \psi^4 $ at $\psi \gg M$, as shown in Fig.\ref{f1}.  We will consider cases where $\langle \psi \rangle$ is close to $M$ which is sufficiently small for field fluctuations to reach the tiny flat region close to $\psi =0$ where large curvature fluctuations can be produced. 
Nevertheless, we will show that this potential cannot simultaneously satisfy the constraints imposed by the CMB anisotropies and the production of PBHs.}

{At the end of the paper, we shortly discuss an other class of potentials, the one of those with an inflection point that can also lead to PBH production without spoiling the large-scale primordial power spectrum.  One could think that most natural example of such scenario is to consider the Brout-Englert-Higgs (BEH) field as the stochastic spectator, given that it exhibits an inflection point for very specific choices of the top quark and the BEH boson mass.  This is the object of a separate paper, where we show that unfortunately this scenario is not viable with our standard knowledge of the BEH potential. }


\section{The Dynamics of the spectator field  after inflation}   
\label{The Dynamics of the Field after inflation}

After the end of inflation, the spectator field remains frozen during the subsequent reheating and radiation era until it dominates the energy density in the Universe, which causes an additional short period of accelerated expansion in some Hubble-sized regions where the field lies in a flat region of the potential.  This results in the generation of large curvature fluctuations, whose statistics will be studied in the next section.   In this section, we focus on the dynamics of the spectator field after inflation.   
 \textcolor{black}{After the field rolls down to the bottom of the potential, it oscillates and possibly lead to a matter-dominated era.  We assume here that it is coupled to other fields such that it quickly decays, driving the Universe back to the radiation era. }

The equations governing the evolution of the spectator field $\mathrm{\psi}$ and of the Universe expansion, in cosmic time, read:
\begin{equation}
\begin{split}
\ddot \psi&+3 H \dot \psi +\frac{\partial V}{\partial \psi} =0~, \\
\dot N&= H =\sqrt{\frac{\rho}{3 M_P^2}} ,
\end{split}
\label{eq:field.t}
\end{equation}
where a dot denotes the derivative {with respect to} the cosmic time.  
 $\mathrm{\rho}$ is the {total energy} density {that evolves as follows}, as long as the field velocity is negligible:
\begin{equation}
\rho =\rho_{\rm m,r} e^{-\kappa N} +V(\psi),
\end{equation}
where $\rho_{\rm m,r}$ is the energy density in matter or radiation at some initial time.    
In case of matter (radiation) domination, one has $\kappa = 3$ ($\kappa = 4$). The initial time must be after inflation but before the field dominates the density of the Universe.  Our choice is  
to consider the time when ${\rho_{\rm r,m}= C \times V(\psi_{\rm ic})}$, with $C = 10$, which typically ensures that the field did not evolve before this time.   We denote the initial field value as $\mathrm{\psi_{ic}}$.

Instead of using the cosmic time, we compute the evolution of the field with respect to the e-fold time.  The Eq.~(\ref{eq:field.t}) can be re-written in efold time in the following way,
\begin{equation}
\psi '' +\psi' \frac{1}{2 \rho} \left(-\kappa \rho_{\rm m,r}e^{- \kappa N}+\psi'\frac{{\rm d} V}{{\rm d} \psi}\right)+3\psi' +\frac{3 
M_{\rm P}^2}{\rho_{\rm m,r}}\frac{{\rm d} V}{{\rm d} \psi}=0~,
\label{eq:field}
\end{equation}
where primes denote derivatives with respect to $N$.   We solve numerically this exact classical equation, instead of assuming the slow-roll approximation as done in~\cite{Carr:2019hud}.  In the rest of the paper, we focus on the case of a radiation phase, $\kappa = 4$.  Our results are nevertheless found to be generic and our conclusions apply to the case $\kappa =3$ as well, with an adequate rescaling of $\langle \psi \rangle$.  



The Eq. (\ref{eq:field}) is numerically solved for a range of initial conditions.  The  numerical integration is stopped when the slow roll parameter $\epsilon_1$ reaches one.  An alternative choice is to continue the integration during the phase where the field oscillates around its minimum, until the density reaches a specific value, e.g. one hundredths of the mean initial density.  As explained in the next section, such a choice would be theoretically more justified given that the $\delta N$ formalism used to calculate the spectrum of curvature fluctuations applies to final hypersurfaces of constant density.  However for the considered scenarios we did not get any appreciable difference between the different choices and the hypersurfaces defined by $\epsilon_1 = 1$ are almost of constant density.  



\begin{figure}[h!]
\centering
\includegraphics[width=80mm]{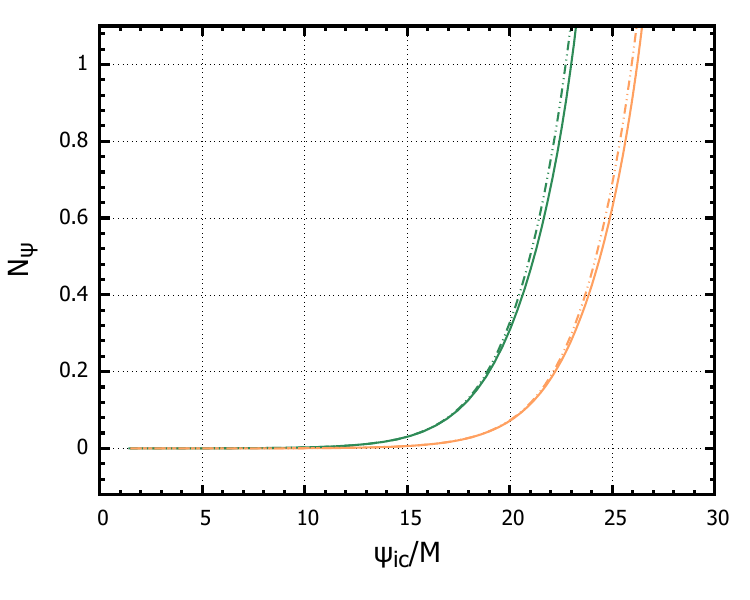}
\caption{ The number of e-folds realized at the time of the spectator field domination, for the  parameters of the potential given in Eq.(\ref{eq:pot1}), as a function of the initial conditions. Solid lines correspond to a radiation dominated era and dashed  lines to a matter era.  }
\label{f3}
\end{figure}

\begin{figure}[h!]
\centering
\includegraphics[width=85mm]{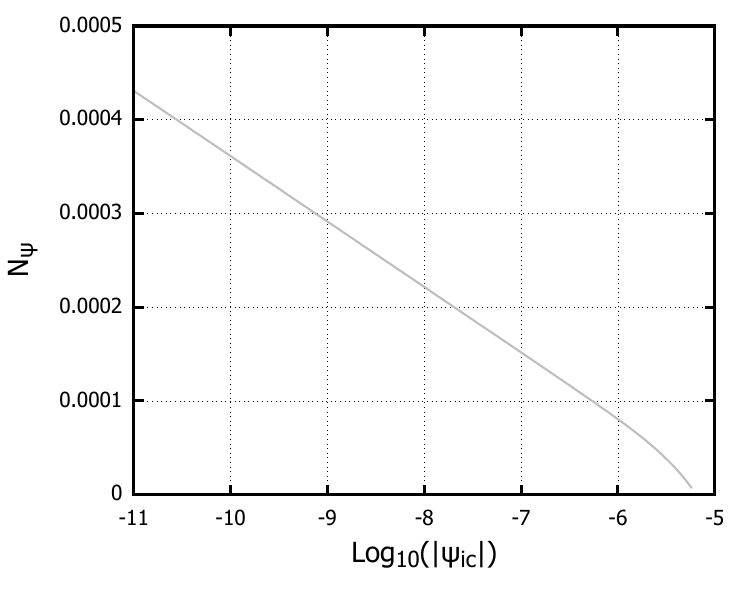}
\caption{  The number of e-folds realized at the time of the spectator field domination, for the  parameters of the potential given in Eq. (\ref{eq:potential_example2}), as a function of the initial conditions 
 and for the parameters given in caption of Fig.\ref{f1}. }
\label{f3b}
\end{figure}




Our results are presented in Fig.\ref{f3} for the two cases $\kappa = 4$ and $\kappa = 3$, for the two considered values of $M$ of the plateau potential (Model 1).  One observes that a brief phase of accelerated expansion is achieved when $\psi_{\rm ic}$ lies in the flat plateau, with a number of generated e-folds that can be larger than one for the considered cases.  This is expected to lead to significant curvature perturbations, later collapsing into PBHs.   For comparison, the results obtained for the small-field potential are displayed in Fig. \ref{f3b} with the same parameter value than in Fig. \ref{f1}.  However, due to the small value of $M$ that we consider -- required to not spoil the primordial power spectrum of CMB scales, as later explained -- one gets that the number of e-folds that are realized always remains very small, much lower than one.   The reason is that even if the potential is extremely flat around $\psi = 0$, the curvature of the potential is large enough to quickly stop the slow-roll phase and drive the field down to the potential minimum.



\section{Production of curvature fluctuations after inflation} 
\label{Production of curvature fluctuations after inflation} 


The fluctuations in the extra expansion can be connected to curvature fluctuations.  Their importance and their statistical distributions can be studied using the stochastic $\mathrm{\delta N}$ formalism.  
It relies on the separate Universe
approximation~\cite{Sugiyama:2012tj} and relates a curvature perturbation $\zeta(x,t)$ on a spatial hypersurface of constant energy density to the  
difference between the number of e-folds of expansion $N(x,t) $ realized starting from an initially flat hypersurface and
the \textit{unperturbed} number of e-folds $\bar N(t)$,
\be
\zeta(x,t) = \delta N^{\rm f}_{\rm i} \equiv N(x,t) - \bar N(t) 
\ee
where we have labelled by $i$ and $f$ the initial and final hypersufaces.  
These are chosen to be at the initial and final time respectively as specified in the previous section.

The curvature fluctuation is a random variable connected to the the random fluctuation of the spectator field value $\delta \psi_{\rm in} $,
i.e. \be
\zeta (x) = N (\psi(x))- \bar N  = N (\delta \psi_{\rm in} + \delta \psi_{\rm out} + \langle \psi \rangle ) - \bar N
\ee
where the mapping between $N$ and $\psi$ comes from the numerical integration of the field trajectories.
The value of $\bar{N}$ and the corresponding $\langle \psi \rangle$ can be arbitrarily chosen, since these are associated to our Hubble-sized patch and result from the stochastic field dynamics during inflation, before observable scales leave the horizon.  In terms of probability distributions, denoted $P$, and using the properties of transformations of random variables, one therefore has for curvature fluctuations associated to a physical size determined by the scale exiting the Hubble radius during inflation at  $N_{\rm inf}$ ,
\be
P(\zeta_{\rm in}-\zeta_{\rm out}) = \int {\rm d} \delta \psi_{\rm out} P(\delta \psi_{\rm in})  P(\delta \psi_{\rm out}) \left. \frac{{\rm d} \psi}{{\rm d} N}\right|_{\psi_{\rm out}} ~. 
\label{eq:probability}
\ee
In this equation, $\psi_{\rm out} = \langle \psi \rangle + \delta \psi_{\rm out}$ and $\delta \psi_{\rm in} = \psi(\zeta_{\rm in}+\zeta_{\rm out}+\langle N \rangle) - \psi_{\rm out}$, i.e. is a function of the curvature fluctuation in the inner region and of the field value in the outer region. Since $\delta \psi_{\rm in} $ and $\delta \psi_{\rm out} $ follow a Gaussian statistics, one has
\begin{equation}
\begin{split} 
    P(\delta \psi_{\rm in})=\frac{1}{\sqrt{2\pi \langle \delta \psi_{\rm in}^2(N_{\rm inf}) \rangle}}  \exp{\left[ \frac{-\delta \psi_{\rm in} ^2 (N_{\rm inf})}{2\langle \delta \psi_{\rm in}^2(N_{\rm inf}) \rangle}\right]} 
    \end{split}
\end{equation}

\begin{eqnarray}
    P(\delta \psi_{\rm out} ) &= &\frac{1}{\sqrt{2\pi \langle \delta \psi_{\rm out}^2(N_{\rm inf}-1) \rangle}}  \nonumber \\
& \times & \exp{\left[ \frac{- \delta \psi_{\rm out}^2 (N_{\rm inf}-1)}{2\langle \delta \psi_{\rm out}^2(N_{\rm inf}-1) \rangle}\right]}~.
\end{eqnarray}
The evolution of the variances $\mathrm{\langle \delta {\psi_{out}}(N_{inf}) \rangle}$ and  $\mathrm{\langle \delta {\psi_{in}}(N_{inf}) \rangle}$ during inflation are given by the relation of Eqs.(\ref{eq:dhout}) and (\ref{eq:dhin}).

In simple words, we consider all the possible realisations of the spectator field in the outside region, weighted by their probability of realisation.  In each of them, we consider the probability to have an inner fluctuation, to which corresponds a variation of the number of e-folds realised, ie. a curvature fluctuation.  It is worth noticing that what is relevant for PBH formation is the curvature fluctuation between the inner region and the outer region and not compared to $\langle N \rangle$, but this subtlety becomes relevant only after a certain number of e-folds, when the values of $\psi_{\rm out}$ can populate the field region where large curvature fluctuations can be generated.  

There is one more condition to be satisfied for our scenario to be viable:  the power spectrum of the curvature fluctuations induced by the stochastic spectator field $\mathcal P_\zeta ^{\psi}$  must not spoil the primordial power spectrum of curvature fluctuations from the inflaton on CMB scales, given by Eq.~(\ref{eq:amplitude}).  In the $\delta N$ formalism applied to our single spectator field, it can be calculated as \cite{Lyth:2005fi,Langlois:2008vk,Sugiyama:2012tj}
\begin{equation}
\mathcal P_\zeta^{\psi} (k)=\frac{H^2 (k)}{4 \pi^2}\left(  \left. \frac{{\rm d}N}{{\rm d} \psi}\right|_{\langle N \rangle } \right)^{2} \ll A_{\rm s}~,
\label{eq:as}
\end{equation}
where $k = k_* \exp (N_{\rm inf}-N_*)$.  
The above mentioned condition can be respected if one selects $\langle \psi \rangle$ to be in the region where $N(\psi)$ is very flat.  For the plateau potential, this is obtained when $\langle \psi \rangle \lesssim M$, and for the small-field potential when it is close to the potential minimum at $\psi = \pm M$.  
The value of $\mathcal P_\zeta^{\psi}$ obtained for the two considered inflationary models given in Table \ref{tab:epsi} and $M$ values as reported in the caption of Fig.\ref{f1} are $\mathrm{ 2.61\times 10^{-10}}$ and $\mathrm{4.97 \times 10^{-11}}$, i.e. in both case below $A_{\rm s}$.  It is also worth noticing that $\mathcal P_\zeta^{\psi}$ typically does not evolve much when considering smaller scale and would still remain below the CMB power spectrum if the statistics of the curvature fluctuations would not develop heavy non-Gaussian tails, as explained hereafter.

\begin{figure}[h!]
\centering
\includegraphics[width=100mm]{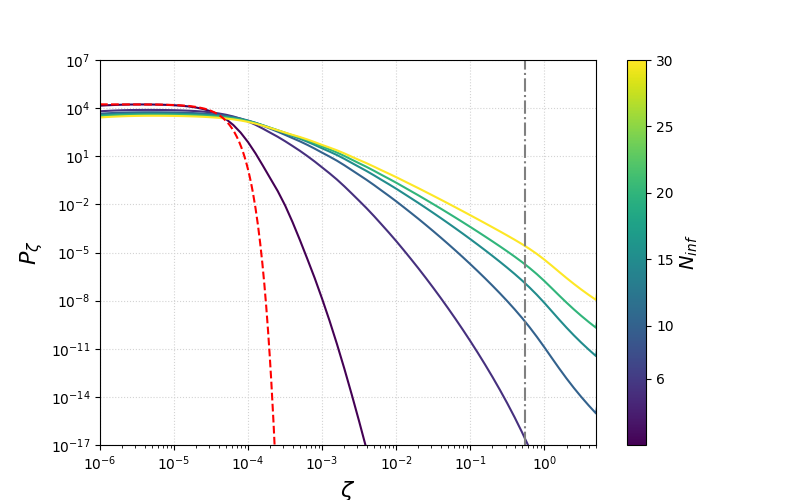}
\caption{ The probability distribution of curvature fluctations $\zeta$ for Model 1 and for different values of $\text{N}_{\text{inf}}$.  The red dashed line corresponds to the case of a Gaussian distribution on CMB scales, determined with Eq.~(\ref{eq:as}).  The vertical gray dash-dotted line denotes the critical threshold $\zeta_{\rm cr} $ for PBH formation. }
\label{f4a}
\end{figure}
{Finally, we proceed to the calculation of the probability distribution of curvature fluctuations, described by Eq. (\ref{eq:probability}).  The outcome is depicted in Fig. \ref{f4a} for the considered Model 1 and for different values of $N_{\rm inf}$.  
We have also shown for comparison 
the Gaussian distribution expected from the calculation of $\mathcal P_\zeta (k)$ only, i.e. neglecting non-Gaussian effects.   On large cosmological scales, i.e. when $N_{\rm inf} < 10$, the Eq. (\ref{eq:probability}) reproduces well the distribution expected from the power spectrum.
One can nevertheless already observe a non-Gaussian tail at $\zeta \gtrsim 10^{-4}$, but not enough to lead to the production of PBHs with a significant probability.   When $N_{\rm inf}$ increases, the tail becomes more and more non-Gaussian and for $N_{\rm inf} \gtrsim 20$ one gets that order one curvature fluctuations are generated, with a very small but non-negligible probability. These will collapse into PBHs when they re-enter inside the horizon.


\section{Primordial Black Hole formation and mass distribution}
\label{Primordial Black Hole formation}

When the curvature fluctuations are generated, they are still super-horizon.   Only later, when they re-enter inside the Hubble radius, they collapse into PBHs when they exceed a certain threshold $\zeta_{\rm cr}$.  One usually denotes $\beta(M_{\rm PBH})$ the density fraction of the Universe collapsing into PBHs of mass $M_{\rm PBH}$ per unit of logarithmic mass interval.  In our scenario, it is therefore obtained by integrating the probability density function of curvature fluctuations above this threshold value,

\begin{equation}
\beta(M_{\rm PBH}) \equiv \frac{1}{\rho}\frac{{\rm d} \rho }{{\rm d} \ln M_{\rm PBH}} = \int_{\zeta_{\rm cr}}^{\infty}  P(\zeta) {\rm d} \zeta \,.
\label{eq:beta_derivative}
\end{equation}

The exact threshold leading to PBH formation depends on several factors, including the equation of state of the Universe which itself depends on the thermal history.  In particular, at the QCD epoch, the slight transient reduction of the equation-of-state slightly reduces the threshold, which can result in a high peak in the PBH mass distribution around $M_{\rm PBH} \approx 2 M_\odot$ and a bump in the range from $30 M_\odot$ to $100 M_\odot$~\cite{Byrnes:2018clq,Carr:2019kxo}.  In order to consider this effect, we have used the overdensity threshold values $\delta_{\rm cr}$ as a function of the Hubble mass at horizon crossing from Ref.~\cite{Escriva:2022bwe}, related to the PBH mass by a factor $\gamma \equiv M_{\rm PBH} / M_H \simeq 0.8  $, recently computed with numerical relativity simulations taking into account the changes in the equation-of-state at the QCD epoch.  One can relate the curvature and density fluctuations through the approximate relation $\zeta \approx (9/4) \delta$ valid at horizon crossing.  The exact relation depends on various factors, such as the shape of the primordial power spectrum, the statistics of the fluctuations (Gaussian or non-Gaussian), the curvature and density profiles \cite{Musco:2023dak,Franciolini:2022tfm,Atal:2019erb}.  All these factors can induce changes in the features visible in the final PBH mass distribution, but without loosing the generality and the viability of our PBH formation scenario.  Then, the dark matter fraction made of PBHs today is obtained by~\cite{Byrnes:2018clq}
\begin{equation}
f_{\rm PBH}(M_{\rm PBH}) \approx 2.4 \beta(M_{\rm PBH}) \left( \frac{2.8 \times 10^{17} M_\odot}{M_{\rm PBH}} \right)^{1/2}~. 
\end{equation} 

The PBH distrubtions at formation and today, $\beta(M_{\rm PBH})$ and $f_{\rm PBH} (M_{\rm PBH})$ respectively, for the Model 1 and Model 2, are diplayed in Figs.~\ref{f6a} and~\ref{f6}.   In both cases, one gets an extended mass distribution with the QCD-induced features clearly visible in the stellar-mass range.  We have selected the parameter $\langle \psi \rangle$ such that one gets a total dark matter density $f_{\rm PBH}^{\rm tot} = 1$, when it is integrated over the full mass range, but different choices can lead to different abundances. 

\textcolor{black}{ By modifying the value of $M$ and of the slow-roll parameters, one influences the mass distribution of PBHs.  A peak around a solar mass is generic, however it is not related to our mechanism but to the variation of the formation threshold at the QCD epoch.  We specifically focus on PBH distributions where this peaks provide most of the abundance in the solar mass range, considering their significant role in observational studies.  But other values of $M $ and inflationary parameters can also generate a decreasing mass distribution dominated by asteroid-mass PBHs, or even lighter PBHs that would have evaporated.   Although this was not the primary focus of our analysis, this illustrates the broad adaptability of our model. }

\begin{figure}[h!]
\centering
\includegraphics[width=80mm]{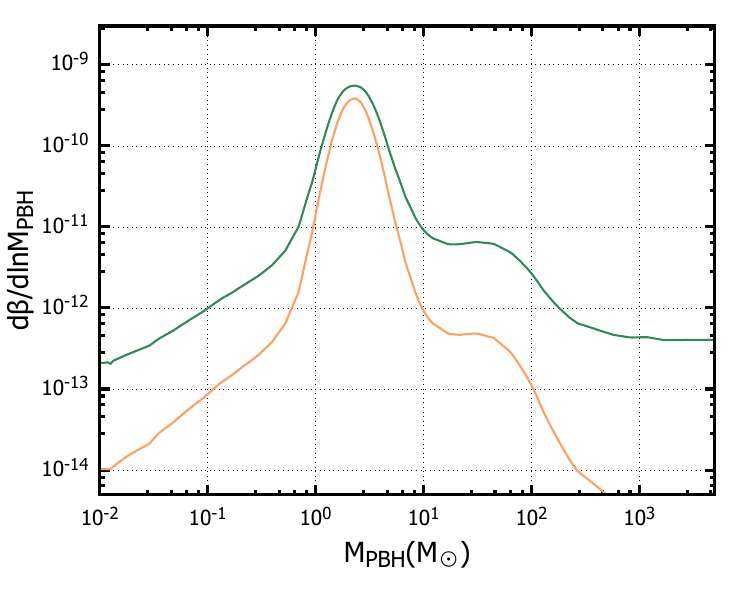}
\caption{ The collapse fraction of PBHs for the two models of inflation: Green line corresponds to Model 1 and orange line to Model 2.} 
\label{f6a}
\end{figure} 
 
\begin{figure}[h!]
\centering
\includegraphics[width=80mm]{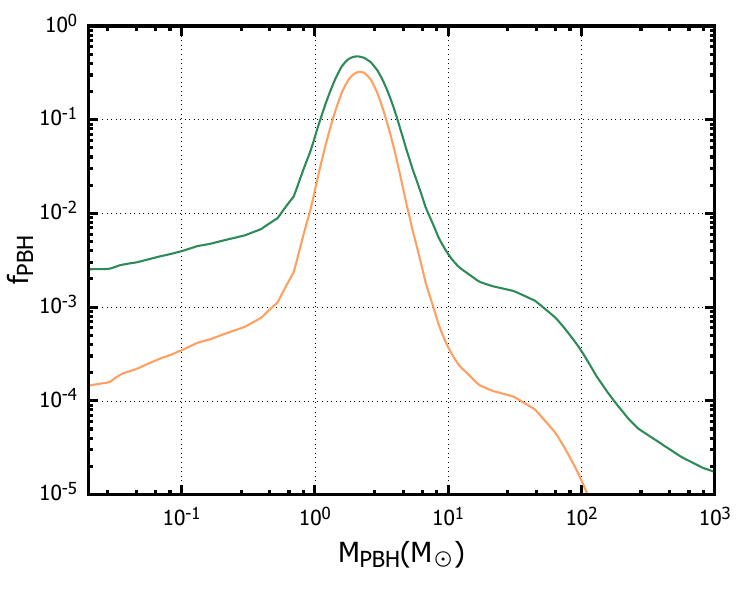}
\caption{ The fraction of PBHs to DM for the two models of inflation: Green line corresponds to Model 1 and orange line to Model 2.} 
\label{f6}
\end{figure}

It is important to emphasize that the probabilities associated with the production of PBHs and their corresponding abundances can be in agreement with the observational constraints of inflation.  With a good choice of $\langle \psi \rangle $ and $M$ one can have at the same time a significant PBH production on small scales and negligible curvature fluctuations on cosmological scales.


\section{Resolving the fine-tuning} \label{fine-tuning}

One of the main criticisms of inflationary models as a mechanism to generate PBHs is the issue of fine-tuning, recently studied in~\cite{Stamou:2021qdk,Cole:2023wyx} for some models. This is in fact a three-sided problem that can be summarized as follows:
\begin{enumerate}
\item The parameters of the inflation model must be fine-tuned in order to produce a high peak in the primordial power spectrum with $\mathcal O(1)$ density fluctuations on on small scales, while not spoiling it on CMB scales where density fluctuations are of $\mathcal O(10^{-5})$.  Adding the constraint that no CMB distortions are produced at an observable level, this particularly restricts the range of possible models~\cite{Byrnes:2018txb,Cole:2022xqc} and further adds up to the required degree of fine-tuning.   
\item In addition to the shape of the primordial power spectrum, requiring that the today abundance of PBHs is sizeable adds another layer of fine-tuning.  This issue is related to the exponential dependence of the PBH abundance with respect to the primordial power spectrum amplitude.  A tiny change of it can typically induce a change in the PBH abundance by several orders of magnitude.  This issue typically increases the PBH fine-tuning problem by one or two orders of magnitude~\cite{Cole:2023wyx}.
\item If the model aims at explaining some GW signals from black hole merger events, one has in addition to explain the coincidence between the PBH mass and stellar masses, and so tune the model parameters to produce a peak in the primordial power spectrum on the correct scales, around $\sim 10^7 \, {\rm Mpc}^{-1}$.  
\end{enumerate}

The last issue can be naturally solved in models leading to a very wide PBH mass distribution. Indeed, the overdensity threshold leading to PBH formation is slightly reduced at the QCD epoch, leading to a peak and a bumpy feature in the PBH mass function in the stellar-mass range, as already mentioned.  The fact that the Hubble mass at the QCD epoch and the Chandrasekhar mass are the same up to an order one numerical factor, whatever is the QCD coupling constant, naturally explains the coincidence between the mass of PBHs and of astrophysical black holes.   This argument applies to our model, as explained in the previous section and as one can see in the obtained PBH mass distributions.  

The first two issues were studied quantitatively for a few representative models in~\cite{Cole:2023wyx}.  One way to quantify the needed fine-tuning on a parameter $p$ to obtain a given observable $\mathcal O$ is to define a measure 
\begin{equation}
\epsilon_{\mathcal O} \equiv \frac{{\rm d} \log \mathcal O}{{\rm d} \log p}~.
\end{equation}
Roughly, this definition is similar to the inverse of the Bayesian evidence of a model with a uniform prior on a unit interval of $\log p$.  The ratio of two values of $\vert \epsilon_{\mathcal O}\vert$ obtained in two different models represents the odds of one with respect to the other.  Let us note that such a measure is not unambiguous, as discussed in details in~\cite{Cole:2023wyx}.   For the few PBH models audited in~\cite{Cole:2023wyx}, very large values of $\vert \epsilon_{\mathcal P_{\rm peak}} \vert $ are obtained, ranging from $10^2$ to $10^8$.  In turn, $\vert \epsilon_{ f_{\rm PBH}} \vert $ ranges from $10^4$ to $10^9$.  On the opposite, an ideal model that would generically lead to $f_{\rm PBH} \sim \mathcal O(1)$ within the same range of $p$ would have a measure $\vert \epsilon_{ f_{\rm PBH}} \vert \sim \mathcal O(1)$ and would be highly favored on a Bayesian statistical point of view. 

In our scenario, the needed fine-tuning is strongly suppressed because of the inevitable stochasticity of the spectator field during inflation, a suppression that we can try to quantify.  For our scenario to work, the only requirement is that the single parameter $M$ of the field potential is smaller but of same order than $H_*$.  But within this restriction, a variation of $M$ can be compensated by a suitable choice of $\langle \psi \rangle$ to obtain any given PBH abundance.  The crucial point is that $\langle \psi \rangle$ is not a model parameter but a stochastic variable that naturally takes a whole possible range of values in the entire Universe.   

Taken alone, however, this argument is insufficient to reduce the fine-tuning issue.  Indeed nothing guarantees one gets $f_{\rm PBH} \sim \mathcal O(1)$ in our observable patch of the Universe, given that many realisations and many values of $f_{\rm PBH}$ are obtained in the different patches.   Obviously $f_{\rm PBH} \gg 1$ can be discarded by invoking an anthropic selection argument:  in those Universe patches, PBH rapidly accrete all the ordinary matter and they become dominated by PBHs only, without galaxies, stars and planets.  But values of $f_{\rm PBH} \ll 1$ are a priori not excluded and a Universe without dark matter is, as-far-as we know, suitable for galaxies and stars to form and for life to appear.   But recently, it was pointed out that PBHs from the QCD epoch can at the same time be at the origin of baryogenesis~\cite{Carr:2019hud,Garcia-Bellido:2019vlf} within the standard model of particle physics, without invoking additional CP violation.   A sizable fraction of the overdensities collapsing into PBH is converted into baryonic matter and therefore there is a connection between the PBH abundance at formation and the baryon-to-photon ratio $\eta \sim \beta \sim 10^{-10}$, as well as a connection between $\Omega_{\rm PBH}$ from the QCD epoch and the baryonic content of the Universe $\Omega_{\rm b}$.   Combining our model to this scenario -- which does not require any additional ingredient or parameter -- one therefore gets a value of the relevant fine-tuning measure $\epsilon_{(f_{\rm PBH} / \Omega{\rm b)}} \sim 1$, given that any value of $M$ in the considered range can lead to $f_{\rm PBH} \sim 1$, while not spoiling CMB observations.   Given that one has to fix the scale of $M$, the evidence of the scenario is in fact reduced by one order of magnitude with  $\epsilon_{(f_{\rm PBH} / \Omega{\rm b)}} \sim \mathcal O(10)$, which cannot be considered as a fine-tuning.  In other words, on a Bayesian point of view, the odds are in favor of our model when compared to the ones studied in~\cite{Cole:2023wyx} by at least one against one thousand.  It could be studied, nevertheless, if for some particularly motivated and less fine-tuned models like critical Higgs inflation~\cite{Ezquiaga:2017fvi,Ezquiaga:2019ftu}, the odds may become comparable.  

\textcolor{black}{The previous discussion does not mention a possible fine-tuning arising from the requirement that quantum correction to the spectator field potential does not spoil the condition that the field is light during inflation, i.e. $\Lambda^4 /M^2\ll H_{\rm *}^2 $.  Such a fine-tuning is for instance present in the Standard Model in order to keep a Brout-Englert-Higgs field mass much below the Planck scale.  }



\section{Discussion and conclusion}
\label{Discussion and conclusion}

PBHs have recently seen a renewed interest as an explanation of GW observations and of the dark matter in the Universe.  However, despite the fact that dozens of formation scenarios have proposed, the vast majority them still suffer from a strong fine-tuning problem.  We have explored a mechanism based on a light stochastic spectator field during inflation, acting as a curvaton and for which the needed parameter tuning is importantly reduced, making PBH formation more natural.  

During inflation, due to its quantum fluctuations, the spectator field explores all its potential.  After inflation, during the subsequent matter or radiation era, there is a time when it starts to dominate the energy density of the Universe.  In the regions of the Universe where the field lies in a flat part of the potential, an extra expansion occurs, corresponding to a curvature fluctuation that can later collapse into a PBH, when it reenters inside the Hubble radius.   Large curvature fluctuations are therefore produced only in rare regions, whereas they remain small in the rest of the Universe.  Compared to Ref.~\cite{Carr:2019hud}  where this scenario was first proposed, we have considerably improved and refined the computation of the field dynamic before and after inflation, the calculation of the statistical distribution of the curvature fluctuations and the resulting PBH abundance.  On large scales, we find that only small, essentially Gaussian, curvature fluctuations are generated and they can be subdominant compared to the inflationary fluctuations in such a way that they do not spoil the primordial power spectrum on CMB scales.  Then, going to smaller scales, the probability distribution develops a heavy non-Gaussian tail, with a small fraction of curvature fluctuations that is above the threshold for PBH formation.   When considering the effects of the QCD epoch on this threshold, one obtain a broad PBH mass distribution covering decades of masses but with a peak between 2 $M_\odot$ and 5 $M_\odot$ and a bumpy feature between 20 $M_\odot$ and 100 $M_\odot$.  Such a distribution is approximately reminiscent to the one expected for a nearly scale invariant primordial power spectrum of Gaussian fluctuations.  But in our scenario the origin and the statistics of curvature fluctuations are radically different.  

We have applied this mechanism to two types of spectator field potential -- plateau and small-field -- and two inflation scenarios.  We have shown that for small-field potentials, it is unfortunately not possible to produce PBHs and have at the same time subdominant curvature fluctuations on cosmological scales, contrary to what was claimed in~\cite{Carr:2019hud}.  This difficulty arises due to the extremely tiny size of the flat region of the potential from which large curvature fluctuations can be produced.   But a plateau potential does the job and we obtained the corresponding PBH mass distributions in this case. 

Even if one does not need a fine-tuning of potential parameters, we nevertheless find that one parameter must be smaller but of the same order of magnitude smaller than the Hubble rate during inflation.  Nevertheless, the abundance of PBHs is fixed by the averaged value of the spectator field in our observable Universe.  Because it is a stochastic variable and not a model parameter, one can invoke an anthropic selection -- different from an antrhopic principle -- to argue that the abundance of PBHs today must be comparable to the one of dark matter.   We also argued that explaining the coincidence between the density of dark matter constituted of PBHs and baryons is eased in a scenario where PBHs are also at the origin of baryogenesis based on the CCGB mechanism also proposed in~\cite{Carr:2019hud}.  This mechanism explains at the same time the observed value of the baryon-to-photon ratio in the Universe that is connected to the density of stellar-mass PBHs produced at the QCD epoch.   

Our present work opens interesting perspectives.  First, the  mass distribution of PBHs could be used to derive PBH merger rates that can be compared to the ones of compact binary coalescences inferred from LVK observations.  This is one way to constrain the exact shape of the potential associated to the spectator field, as well as the underlying inflationary dynamics, in particular the energy scale of inflation.  Second, one could investigate the effect of the fully non-Gaussian distribution of curvature fluctuations onto the scalar-induced GW spectrum.  The recent PTA observations of a GW background at nano-Hertz frequencies suggest that the signal is hardly compatible with models of PBH formation from Gaussian fluctuations.  But the signal could be explained by PBHs coming from non-Gaussian fluctuations~\cite{Firouzjahi:2023xke}.  Third, one may consider to explore scenarios based on the same mechanism and embedded in existing high-energy frameworks, for instance models based on no-scale supergravity.  The most realistic candidate could be the Brought-Englert-Higgs field itself, but only if its potential has a flat region or an inflexion point at field values comparable to the Hubble expansion during inflation, due to radiative corrections, but which may require a non-minimal coupling to gravity or deviations from the standard model predictions.   Finally, we envision that some scenarios may exist where the large-scale CMB fluctuations arise from the stochastic spectator field itself, as originally proposed in curvaton scenarios, while PBHs are produced on smaller scales from the non-Gaussian tails.   

Overall, we pave the road to new analysis and new models that would address the critical fine-tuning issue related to PBHs, while possibly explaining puzzling observations, the most important one being the existence of the dark matter in the Universe.

%


\bibliographystyle{apsrev4-2}

\bibliography{bib.bib} 




\end{document}